\documentclass[12pt]{article}
\usepackage{booktabs}  
\usepackage{amsmath}
\usepackage{amsthm}
\usepackage{amssymb}
\usepackage{mathrsfs}
\usepackage{gensymb}
\usepackage{textcomp}  
\usepackage{xcolor}
\usepackage{lipsum}    
\usepackage{graphicx}
\usepackage[font=footnotesize]{caption}
\usepackage{subcaption}
\usepackage{bm, bbm}
\RequirePackage{diagbox, adjustbox}
\usepackage[colorlinks=true,linkcolor=black, citecolor=black, urlcolor=black]{hyperref}
\usepackage{multirow}
\usepackage{multicol}
\usepackage{pdflscape}
\usepackage{afterpage}
\usepackage{everypage}
\usepackage{natbib}
\usepackage{float}
\usepackage{url} 

\begin{document}

\def\spacingset#1{\renewcommand{\baselinestretch}%
{#1}\small\normalsize} \spacingset{1}


\begin{center}
{\Large { \bf An Adaptive Graphical Lasso Approach to Modeling Symptom Networks of Common Mental Disorders in Eritrean Refugee Population}}
\end{center}

\vspace{.05in} 

\begin{center}
{\large { Elizabeth B. Amona$^{1}$, Indranil Sahoo$^{{1}*}$, David Chan$^{2}$, Marianne B. Lund$^{3}$, Miriam Kuttikat$^{3}$}} \\

\bigskip

{\normalsize {\it $^{1}$Department of Statistical Sciences and Operations Research, Virginia Commonwealth University, Richmond, VA, USA \\
$^{2}$Department of Mathematics and Applied Mathematics, Virginia Commonwealth University, Richmond, VA, USA \\ 
$^{3}$School of Social Work, Virginia Commonwealth University, Richmond, VA, USA \\
$^{4}$Biostatistics Core, Brown Cancer Center, University of Louisville,  Louisville, KY, USA \\}}

\vspace{.1in}

$^*$Corresponding author; Email: \url{sahooi@vcu.edu}\\
\end{center}

\vspace{.1in}

\vspace{.0075in}
\baselineskip 18truept

\bigskip
\begin{abstract}
 
\noindent Despite the significant public health burden of common mental disorders (CMDs) among refugee populations, their underlying symptom structures remain underexplored. This study uses Gaussian graphical modeling to examine the symptom network of post-traumatic stress disorder (PTSD), depression, anxiety, and somatic distress among Eritrean refugees in the Greater Washington, DC area. Given the small sample size ($n$) and high-dimensional symptom space ($p$), we propose a novel extension of the standard graphical LASSO by incorporating adaptive penalization, which improves sparsity selection and network estimation stability under $n < p$ conditions. To evaluate the reliability of the network, we apply bootstrap resampling and use centrality measures to identify the most influential symptoms. Our analysis identifies six distinct symptom clusters, with somatic-anxiety symptoms forming the most interconnected group. Notably, symptoms such as nausea and reliving past experiences emerge as central symptoms linking PTSD, anxiety, depression, and somatic distress. Additionally, we identify symptoms like feeling fearful, sleep problems, and loss of interest in activities as key symptoms, either being closely positioned to many others or acting as important bridges that help maintain the overall network connectivity, thereby highlighting their potential importance as possible intervention targets.

\end{abstract}

\noindent%
 

\section{Introduction}
The global prevalence of Common Mental Disorders (CMDs), including depression, anxiety, post-traumatic stress disorder (PTSD), and somatic symptoms has significant impact on individual and public health \citep{byrow2019barriers, byrow2020perceptions,byrow2022profiles, silove1997anxiety}. Refugees and asylum seekers are particularly vulnerable to these disorders due to exposure to traumatic events before, during, and after migration (see \cite{berhe2021prevalence}). Among migrant populations, particularly those from conflict-affected regions like Eritrea, the burden of CMDs is often exacerbated by cumulative stressors across multiple stages of migration. These include pre-migration trauma, perilous journeys, and post-migration challenges such as legal insecurity, unemployment, housing instability, discrimination and intergenerational conflict \citep{chu2013effects, miller2017mental, sengoelge2022post}. Such adversities are well-documented contributors to psychological distress in similarly displaced populations \citep{portes2001legacies, sahoo2024enhancing}, yet Eritrean migrants remain largely underrepresented in mental health research, limiting the development of culturally responsive interventions tailored to their needs \citep{chernet2021mental}.

Eritrean migrants, fleeing political violence, forced conscription, and economic hardship, often rely on tight-knit social networks for coping, resource access, and resilience-building. Although Eritrean immigrants constitute a relatively small proportion of the U.S. population (approximately 35,900 as of 2016), they are highly concentrated in states such as California, Texas, and Washington. Within these communities, traditional epidemiological approaches often fail to capture the complex, interrelated nature of CMD symptoms, necessitating advanced methodologies that explicitly model symptom interdependencies. 

Prior research has identified several key CMD symptoms such as trouble staying asleep, feeling distant, feeling upset, disturbing memories, hypervigilance and somatic symptoms such as nausea or upset stomach and hot/cold spells \citep{Lies2021, Duek2021, Zhou2020, Bryant2017, Schumacher2023}. However, these studies primarily relied on traditional diagnostic models, which treated mental health conditions as isolated constructs and assumed that symptoms operated independently rather than as part of an interconnected network. As a result, these studies failed to capture the complex ways in which symptoms reinforce each other, leading to an incomplete understanding of CMD symptom dynamics.

In this paper, we model the standardized CMD symptoms network among Eritrean migrants in the Greater Washington, DC area using a Gaussian graphical model \citep[GGM; ][]{dobra2004sparse}. Under the assumption of multivariate normality of standardized ordinal symptom scores, the GGM-based network model uses the inverse of the symptom covariance matrix as a measure for symptom associations. Here, we shift from traditional diagnostic frameworks to a symptom-level perspective \citep{briganti2024network}, which allows us to identify central symptoms that drive persistence and co-occurrence, analyze symptom clustering across CMD domains, and assess network stability through different metrics \citep{epskamp2018estimating}. By mapping symptom connections, we highlight the central features that reinforce psychopathology, recognizing that mental health symptoms interact dynamically rather than functioning in isolation.

A key challenge in our analysis is the small sample size ($n$) relative to the high-dimensional symptom space ($p$). To address this, we extended the standard Graphical LASSO framework by incorporating adaptive penalization \citep{zou2006adaptive}, resulting in a novel application we refer to as Adaptive Graphical LASSO for CMD symptom networks. This adaptive approach enhances sparsity selection and improves stability under $n < p$ conditions \citep{fan2009network}. Additionally, we follow the recommendations of \cite{epskamp2018estimating} to evaluate the accuracy of the estimated network parameters, further strengthening the robustness of our findings. While the standard Graphical LASSO \citep[\textit{glasso};][]{friedman2008sparse} has been widely applied in biological studies \citep{huang2020application, poikonen2022graphical}, our study introduces a novel adaptive extension tailored to modeling psychological symptom networks under high-dimensional conditions, designed to enhance sparsity selection and network stability in this setting. By applying this framework to Eritrean migrants, we capture a systems-level understanding of the psychological burden within this underrepresented community, shaped by cultural, structural, and migration-related stressors, thereby informing targeted interventions.

This work is organized in the following manner.  Section~\ref{sec:Data} presents the Eritrean dataset utilized in this study. Section~\ref{sec:Method} outlines the methodology. The results are provided in Section~\ref{sec:Result}. Finally, Section~\ref{sec:Discuss} offers a discussion of the method and results, along with recommendations for practical implementation.

\section{Eritrean Mental Health Data}
\label{sec:Data}

This study utilizes data collected from $n = 19$ Eritrean migrants residing in the Greater Washington, DC area through structured surveys and clinical mental health assessments. Participants were recruited via community organizations, refugee assistance programs, and snowball sampling \citep{johnson2014snowball}. Eligibility criteria included Eritrean descent, age 18 years or older, and a minimum residency of 18 years in the United States. Surveys were administered in both English and Tigrinya by trained bilingual facilitators to ensure comprehension. Additionally, data on demographic variables such as age, living conditions, family structure, economic status, and resource utilization were collected to contextualize mental health outcomes within participants' broader social environments. The dataset consists of 41 symptoms and associated factors related to Common Mental Disorders (CMDs). These variables are categorized into four domains: Somatic Symptoms, Anxiety Symptoms, Depression Symptoms, and PTSD Symptoms, with each variable representing a specific symptom or behavioral indicator in the network model (see Table \ref{tab:descriptive_stats}).

Mental health outcomes were assessed using validated clinical scales widely applied in refugee and migrant mental health research \citep{sahoo2024enhancing, kuttikat2025conceptual}. Depression, anxiety, somatization, and hostility were measured using the Brief Symptom Inventory \citep[BSI-53;][]{derogatis1993brief}, while PTSD symptom severity was assessed using the Posttraumatic Stress Checklist \citep[PCL-5;][]{weathers1993ptsd}. Functional impairment was evaluated using the World Health Organization Disability Assessment Schedule \citep[WHODAS-II; ][]{ustun2010measuring}, and daily stressors specific to migrant populations were measured via the Afghan Daily Stressors Scale \citep[ADSS; ][]{miller2008daily}. Adolescent psychological well-being was assessed with the Strengths and Difficulties Questionnaire \citep[SDQ; ][]{goodman1997strengths}. Additionally, measures assessing mental health literacy, stigma-related attitudes, and help-seeking behaviors were included to provide insight into participants’ knowledge and perceptions of mental health services. All study procedures were approved by the authors' University's Institutional Review Board (\#HM20018353). Also, informed consent was obtained from all participants, with each signing a consent form prior to data collection.

\subsection{Exploratory Data Analysis}\label{demo}
The data consisted of Eritrean migrants aged between 45 and 71 years, with a mean age of 59.58 years (SD = 7.52). Household composition varied, with family sizes ranging from 2 to 7 members and an average family size of 4 (SD = 2). Living conditions, assessed on a scale of 3 to 9, had a mean score of 5.32 (SD = 1.83), indicating moderate variability in housing stability among participants. Regarding education levels, 52.6\% of participants had completed elementary school, 21.1\% attained a high school or GED qualification, 15.8\% had a college/university education, 5.3\% attended nursery school, and 5.3\% pursued advanced graduate or professional education. 

Participants reported varying levels of CMD symptoms across somatic, anxiety, depression, and PTSD domains. Somatic symptoms, ranging from fainting/dizziness to feeling weak in the body, had a mean total score of 10.11 (SD = 4.23; range: 0 – 17). Anxiety symptoms, spanning from pre-migration anxiety to a sense of terror and panic, had a mean total score of 15.43 (SD = 5.42; range: 8 – 23). Depression symptoms, covering experiences from pre-migration depression to thoughts of ending one’s life, had an average total score of 38.05 (SD = 10.52; range: 17 – 53). PTSD symptoms, ranging from pre-migration PTSD to hypervigilance and exaggerated startle response', had the highest overall mean score of 46.16 (SD = 12.99; range: 17 – 62). Here, symptom level means and standard deviations were calculated for standardized symptom scores across all individuals. A detailed breakdown of individual symptom severity and distribution is presented in Table \ref{tab:descriptive_stats}. In addition, examining the covariance structure across symptom scores revealed high sample correlations between multiple pairs of symptoms, such as feeling upset and avoiding activities, and experiencing terror and difficulty breathing, thereby offering an initial view into how symptoms are interrelated.

\begin{table}[]
\centering
\caption{Descriptive statistics (Means, SD and Range) and variable descriptions for demographic characteristics and common mental disorder (CMD) symptoms among Eritrean migrants ($n = 19$). For both demographic and symptom variables, means and standard deviations were computed across individuals. }
\label{tab:descriptive_stats}
\resizebox{\textwidth}{!}{%
\begin{tabular}{l l c c c}
\hline
\textbf{Category} & \textbf{Variable Name} & \textbf{Mean (SD)} & \textbf{Range} & \textbf{Description} \\  
\hline
\multicolumn{5}{l}{\textbf{Demographics}} \\
Age (years) & -- & 59.58 (7.52) & 45 – 71 & -- \\ 
Family Size & -- & 4.11 (1.52) & 2 – 7 & -- \\ 
Living Condition & -- & 5.32 (1.83) & 3 – 9 & -- \\ 
Education Level & -- & -- & See Subsection \ref{demo} & -- \\ 
\hline
\multicolumn{5}{l}{\textbf{Somatic Symptoms}} \\
Som1 & Fainting/Dizziness & 1.68 (0.75) & 0 – 3 & Fainting or dizziness \\ 
Som2 & Chest Pain & 1.58 (0.84) & 0 – 3 & Chest pain \\  
Som3 & Nausea & 1.11 (0.74) & 0 – 3 & Nausea or upset stomach \\  
Som4 & Trouble Breathing & 1.74 (0.73) & 0 – 3 & Difficulty breathing \\  
Som5 & Hot/Cold Spells & 1.0 (0.67) & 0 – 2 & Experiencing hot or cold spells \\  
Som6 & Numbness/Tingling & 1.05 (0.62) & 0 – 2 & Numbness or tingling in body \\ 
Som7 & Weakness in Body & 1.95 (0.91) & 0 – 3 & Feeling weak in body \\ 
SomTot & Total Somatic Symptoms & 10.11 (4.23) & 0 – 17 & Sum of all somatic symptoms \\  
\hline
\multicolumn{5}{l}{\textbf{Anxiety Symptoms}} \\
Anx1 & Anxiety Prior to U.S. & 1.37 (0.83) & 1 – 4 & Anxiety level before migration \\ 
Anx2 & Anxiety Since U.S. & 2.68 (1.1) & 1 – 5 & Anxiety level after migration \\ 
Anx3 & Nervousness & 2.53 (1.07) & 1 – 4 & Feeling nervous \\ 
Anx4 & Tension & 3.0 (0.94) & 1 – 4 & Feeling tense \\ 
Anx5 & Sudden Fear & 1.32 (0.58) & 1 – 3 & Experiencing sudden fear \\ 
Anx6 & Fearfulness & 3.16 (0.96) & 1 – 4 & Feeling fearful \\ 
Anx7 & Sense of Terror & 2.74 (0.99) & 1 – 4 & Sense of terror or panic \\ 
AnxTot & Total Anxiety Symptoms & 15.43 (5.42) & 8 – 23 & Sum of all anxiety symptoms \\  
\hline
\multicolumn{5}{l}{\textbf{Depression Symptoms}} \\
Dep1 & Depression Prior to U.S. & 1.37 (0.68) & 1 – 3 & Depression before migration \\ 
Dep2 & Depression Since U.S. & 1.95 (0.96) & 1 – 4 & Depression after migration \\ 
Dep3 & Feeling Blue & 2.05 (0.97) & 1 – 4 & Feeling sad or down \\ 
Dep4 & Loss of Interest & 2.03 (1.01) & 1 – 4 & Lack of interest in activities \\ 
Dep5 & Hopelessness & 1.26 (0.73) & 1 – 4 & Feeling hopeless \\ 
Dep6 & Worthlessness & 1.42 (0.51) & 1 – 2 & Feeling worthless \\ 
Dep7 & Loneliness & 2.11 (0.99) & 1 – 5 & Feeling lonely \\ 
Dep8 & Suicidal Thoughts & 1.63 (0.68) & 1 – 3 & Thoughts of ending life \\ 
DepTot & Total Depression Symptoms & 38.05 (10.52) & 17 – 53 & Sum of all depression symptoms \\  
\hline
\multicolumn{5}{l}{\textbf{PTSD Symptoms}} \\
PTSD1 & PTSD Prior to U.S. & 1.26 (0.65) & 1 – 3 & PTSD level before migration \\  
PTSD2 & PTSD Since U.S. & 1.89 (0.81) & 0 – 4 & PTSD level after migration \\ 
PTSD3 & Disturbing Memories & 3.0 (0.82) & 1 – 4 & Intrusive memories or thoughts \\ 
PTSD4 & Disturbing Dreams & 2.74 (0.99) & 1 – 4 & Disturbing dreams of stressful experiences \\  
PTSD5 & Reliving Experience & 2.26 (1.15) & 1 – 4 & Feeling as if reliving past experiences \\  
PTSD6 & Feeling Upset & 2.63 (1.02) & 1 – 4 & Feeling very upset \\   
PTSD7 & Physical Reactions & 2.53 (1.02) & 1 – 4  & Having physical negative reactions \\ 
PTSD8 & Avoiding Thoughts & 2.26 (1.05) & 1 – 4  & Avoiding thinking or talking about the stressful past \\
PTSD9 & Avoiding Activities & 2.42 (1.07) & 1 – 5  & Avoiding activities or situations that remind you of the past \\
PTSD10 & Trouble Remembering & 2.53 (1.12) & 1 – 4  & Trouble remembering important parts of stressful experiences \\
PTSD11 & Loss of Interest & 2.68 (0.95) & 1 – 4  & Loss of interest in the things you used to enjoy \\ 
PTSD12 & Feeling Distant & 3.0 (1.15) & 1 – 4 & Feeling cut off from others \\  
PTSD13 & Emotional Numbness & 2.58 (1.02) & 1 – 4  &  Feeling emotionally numb \\ 
PTSD14 & Future Feels Cut Short & 3.16 (1.01) & 1 – 5  & Feeling as if your future will somehow be cut short \\ 
PTSD15 & Sleep Problems & 3.26 (1.45) & 1 – 5 & Trouble staying asleep \\  
PTSD16 & Irritability & 2.89 (1.1) & 1 – 4  & Feeling irritable \\ 
PTSD17 & Trouble Concentrating (PTSD17) & 3.0 (1.05) & 1 – 4  & Trouble concentrating  \\ 
PTSD18 & Hypervigilance & 2.53 (0.96) & 1 – 4 & Being overly alert \\ 
PTSD19 & Easily Startled (PTSD19) & 2.68 (0.82) & 1 – 4  & Feeling jumpy \\ 
PTSDTot & Total PTSD Symptoms & 46.16 (12.99) & 17 – 62 & Sum of all PTSD symptoms \\  
\hline
\end{tabular}%
}
\end{table}

\section{Statistical Methodology} \label{sec:Method}

Let $\mathbf{X} \in \mathbb{R}^{n \times p}$ be the standardized symptom score matrix, where $\mathbf{X}_i = (X_{i1}, X_{i2}, \dots, X_{ip})^\top$ denotes the set of $p$ standardized scores for the $i^{th}$ individual, $i = 1, 2, \ldots, n$. Here, each variable has been standardized separately over $n$ individuals. Prior research suggests that z-score standardization makes ordinal symptom scores approximately normally distributed \citep{epskamp2018estimating}. Let $\mathbf{S}$ denote the $p \times p$ sample covariance matrix for the symptoms, which serves as an empirical estimator of the population-level covariance matrix $\bm{\Sigma}$. 

In high-dimensional settings where $n < p$, the sample covariance matrix is often rank-deficient, which makes direct inversion infeasible and leads to instability in downstream statistical analyses \citep{newman2006modularity, belsley2005regression}. Moreover, small sample sizes introduce high variance in $\mathbf{S} $, leading to poor inference. To ensure numerical stability, as determined by the condition number \citep{cheney1998numerical, belsley2005regression} of the covariance matrix, a ridge regularization term \citep{hoerl1970ridge, hilt1977ridge, gruber2017improving} is introduced. This adjustment ensures a well-conditioned precision matrix even when $n < p$, which captures conditional dependencies between symptom variables and provides the basis for network modeling.


\subsection{Graphical Lasso with Adaptive Regularization}\label{adapt}

Given the sample covariance matrix $\mathbf{S}$, a sparse estimate of the precision matrix $\boldsymbol{\Theta}$ can be obtained using the standard graphical LASSO (\textit{glasso}) algorithm \citep{friedman2008sparse} as follows: 
\begin{equation*}
\widehat{\boldsymbol{\Theta}} = \underset{\boldsymbol{\Theta} \succ \mathbf{0}}{\arg\min} \left( \text{tr}(\mathbf{S} \boldsymbol{\Theta}) - \log \det (\boldsymbol{\Theta}) + \lambda \sum_{i \neq j} |\boldsymbol{\Theta}_{ij}| \right),
\end{equation*}
where $\lambda$ is a regularization parameter controlling sparsity. However, in high-dimensional settings where $n < p$, obtaining a stable estimate of the precision matrix using the \textit{glasso} becomes challenging due to uniform penalization applied across all precision matrix elements. The adaptive graphical LASSO addresses this limitation by incorporating adaptive penalty weights, which allows for differential shrinkage that better preserves important relationships while still enforcing sparsity. Let 
\begin{equation*}
\omega_{ij} = \frac{1}{|\widehat{\boldsymbol{\Theta}}_{\text{init}, ij}| + \delta}
\end{equation*}
denote the element-wise penalties in the adaptive \textit{glasso} algorithm, where $\widehat{\boldsymbol{\Theta}}_{\text{init}}$ denotes the initial \textit{glasso} estimate, and $\delta = 0.2$ prevents division by zero. The precision matrix is then estimated via:
\begin{equation*}
\widehat{\boldsymbol{\Theta}}_{\text{adaptive}} = \underset{\boldsymbol{\Theta} \succ 0}{\arg\min} \left( \text{tr}(\mathbf{S} \boldsymbol{\Theta}) - \log \det (\boldsymbol{\Theta}) + \lambda \sum_{i \neq j} \omega_{ij} |\boldsymbol{\Theta}_{ij}| \right).
\end{equation*}


To select the optimal regularization parameter \( \lambda \), we perform 5-fold cross-validation by minimizing the average negative log-likelihood of the estimated precision matrix on held-out data. Let \( \mathbf{X} \in \mathbb{R}^{n \times p} \) denote the standardized data matrix used for network estimation. For cross-validation, we partition the rows of \( \mathbf{X} \) into \( K = 5 \) disjoint folds, denoted \( \{\mathbf{X}^{(1)}, \dots, \mathbf{X}^{(K)}\} \), where each \(\mathbf{X}^{(k)} \) contains a subset of observations corresponding to fold \( k \in \{1, \dots, K\} \). In each iteration, the \( k \)-th fold serves as the validation set, and the remaining \( K - 1 \) folds form the training data. Let \( \mathbf{S}^{(k)} \) denote the sample covariance matrix computed from the \( k \)-th validation fold. For each candidate \( \lambda \), we estimate the corresponding precision matrix \( \widehat{\boldsymbol{\Theta}}^{(-k)}(\lambda) \in \mathbb{R}^{p \times p} \), where \( \widehat{\boldsymbol{\Theta}}^{(-k)} \) is the inverse of the covariance matrix estimated from the training folds excluding fold \( k \), and compute the log-likelihood loss for fold \( k \) as:
\[
\boldsymbol{\ell}_k(\lambda) = \text{tr}\left(\mathbf{S}^{(k)} \widehat{\boldsymbol{\Theta}}^{(-k)}(\lambda)\right) - \log \det\left(\widehat{\boldsymbol{\Theta}}^{(-k)}(\lambda)\right),
\]
The overall cross-validation error is:
\[
\text{CV}(\lambda) = \frac{1}{K} \sum_{k=1}^{K} \ell_k(\lambda).
\]

The optimal tuning parameter \( \lambda^* \) is selected as the value minimizing \( \text{CV}(\lambda) \). This criterion is consistent with the penalized log-likelihood framework proposed in \citet{friedman2008sparse} for graphical lasso and adapted to cross-validation in \citet{galloway2018cvglasso}. In our analysis, this procedure was implemented using the \texttt{CVglasso()} function from the R package \texttt{CVglasso}, which performs 5-fold cross-validation over a logarithmic grid of \( \lambda \) values with default \texttt{lam.min.ratio = 0.01}. See \cite{friedman2008sparse} for the graphical lasso formulation and \cite{zou2006adaptive,huang2008adaptive} for a general treatment of cross-validation in the adaptive lasso framework.

The estimated precision matrix defines the underlying conditional dependence structure of the symptoms variables. To construct a graph representation, we define the graph adjacency matrix $\mathbf{W}$ as:
\begin{equation*}
W_{ij} =
\begin{cases}
|\widehat{\boldsymbol{\Theta}}_{\text{adaptive}, ij}|, & \text{if } \widehat{\boldsymbol{\Theta}}_{\text{adaptive}, ij} \neq 0, \\
0, & \text{otherwise}.
\end{cases}
\end{equation*}
with $W_{ii}$ set to 0 to remove self-loops.

\subsection{Community Detection in Symptoms Network}\label{comm}

Next, we perform community detection in the adaptive graphical lasso-estimated symptom network, constructed from the non-zero entries of the precision matrix, to identify patterns of symptom co-occurrence and clusters of symptoms that may share underlying mechanisms. For this purpose, several methods exist in the literature \citep{girvan2002community, blondel2008fast, raghavan2007near, rosvall2008maps}. Here, we use the Walktrap algorithm \citep{pons2005computing} for community detection due to its ability to uncover latent structures in complex networks. The Walktrap algorithm is built on the concept of random walks. Recent research has shown that Walktrap performs well in identifying psychometric and clinical symptom clusters \citep{brusco2022comparison, jamison2021optimizing}, and gives accurate results when combined with the \textit{glasso} algorithm \citep{christensen2024comparing}.

Formally, we represent the network as an undirected graph \( G = (V, E) \), where \( V \) is the set of nodes and \( E \) is the set of edges. The \textit{weighted adjacency matrix} \( \mathbf{W} \) is defined such that \( W_{ij} > 0 \) if an edge exists between nodes \( i \) and \( j \), and \( W_{ij} = 0 \) otherwise. The weights \( W_{ij} \) represent the absolute values of the non-zero entries in the adaptive precision matrix, as defined in Subsection~\ref{adapt}. The probability of transitioning from node \( i \) to node \( j \) in a single step of a random walk is then given by:

\[
P_{ij} = \frac{W_{ij}}{\sum_{k} W_{ik}}, \quad i, j \in V.
\]

Using the \( t \)-step transition probability matrix \( \mathbf{P}^t \), the Walktrap algorithm estimates node distances based on how frequently random walkers visit similar locations \citep{pons2005computing}. The distance between two nodes \( i \) and \( j \) is computed as:

\[
d(i, j) = \sqrt{\sum_{k \in V} \frac{(P^t_{ik} - P^t_{jk})^2}{d_k}},
\]
where \( d_k = \sum_{j} W_{kj} \) represents the weighted degree of node \( k \). Nodes with similar diffusion behavior are grouped using an agglomerative hierarchical clustering approach, optimizing the modularity function \( Q \) to determine the best partitioning:

\[
Q = \frac{1}{2m} \sum_{i,j} \left( W_{ij} - \frac{d_i d_j}{2m} \right) \delta(c_i, c_j),
\]
where \( m = \frac{1}{2} \sum_{i,j} W_{ij} \) is the total sum of edge weights, \( c_i \) represents the community label assigned to node \( i \) by the Walktrap algorithm, and \( \delta(c_i, c_j) \) is an indicator function that equals 1 if \( i \) and \( j \) are in the same community and 0, otherwise. This modularity function was introduced in \cite{newman2004finding} and remains widely used in community detection.

Walktrap’s ability to model both local and global structure makes it well-suited for CMD-related symptom clustering. By leveraging random walks rather than solely relying on modularity optimization, it captures deeper relationships that static methods may overlook. Given its demonstrated success in psychological network research and its strong empirical performance with \textit{glasso}, it serves as a robust approach for our analysis. The Walktrap algorithm is implemented using the \texttt{igraph} package in \texttt{R} \citep{csardi2006igraph}. For network visualization, we use the \texttt{qgraph} package \citep{epskamp2012qgraph}.


\subsection{Node Centrality Measures}
We examine three widely used centrality measures, namely,  \textit{Strength}, \textit{Betweenness}, and \textit{Closeness} to quantify the importance or influence of individual nodes within the estimated network. \textit{Strength} centrality captures the total weight of edges connected to a node, reflecting its level of connectivity and integration within the network \citep{freeman1977set, opsahl2010node}. \textit{Betweenness} centrality identifies nodes that act as intermediaries or bridges along the shortest paths between other nodes, highlighting their role in facilitating communication and interaction across the network \citep{freeman1977set}. \textit{Closeness} centrality measures how efficiently a node can access all other nodes, emphasizing its capacity to disseminate or receive information quickly \citep{sabidussi1966centrality, bavelas1950communication}.

Mathematically, we define the node centrality measures of node $i$ in the estimated network as follows:
\begin{itemize}
    \item \textbf{Strength Centrality}:
    \begin{equation}
        S(i) = \sum_{j \in \mathcal{N}(i)} W_{ij},
    \end{equation}
    where $W_{ij}$ represents the estimated weight of the edge between node $i$ and node $j$, and $\mathcal{N}(i)$ denotes the set of neighbors of node $i$.

    \item \textbf{Betweenness Centrality}:
    \begin{equation}
        B(i) = \sum_{r \neq i \neq t} \frac{\sigma_{rt}(i)}{\sigma_{rt}},
    \end{equation}
    where $\sigma_{rt}$ is the total number of shortest paths between nodes $r$ and $t$, and $\sigma_{rt}(i)$ represents the number of those paths that pass through node $i$. Due to the unstable nature of betweenness centrality and the lack of interpretability and reliability of its indices \citep{epskamp2018estimating}, we specifically use weighted betweenness \citep{medeiros2016weighted} in this work, where the weights are defined as the inverse of the absolute values of the non-zero entries in the estimated precision matrix $\hat{\Theta}_{\text{adaptive}}$.

    \item \textbf{Closeness Centrality}: 
    \begin{equation}
        C(i) = \frac{1}{\sum_{j \in V} s(i, j)},
    \end{equation}
    where $s(i, j)$ represents the shortest path distance between node $i$ and node $j$.
\end{itemize}

\subsection{Bootstrap Analysis for Network Stability}

We conduct a nonparametric bootstrap analysis with \( B = 1000 \) iterations to assess the stability of node centrality estimates under data resampling. For each iteration \( b \in \{1, 2, \dots, B\} \), a bootstrap sample is generated by resampling the rows of the data matrix \( \mathbf{X} \) with replacement. The resampled data are standardized using z-score normalization, and the covariance matrix is computed from the standardized data. A graphical LASSO model is then fit using a cross-validated penalty parameter to obtain an initial estimate of the precision matrix. Based on this estimate, we compute adaptive weights and apply the adaptive graphical LASSO model to obtain the final precision matrix \( \hat{\boldsymbol{\Theta}}_{\text{adaptive}}^{(b)} \). We construct a network \( G^{(b)} \) from the nonzero entries of the estimated precision matrix and calculate the centrality measure for each node.

This procedure yields bootstrap distributions of centrality values, which we use to evaluate the consistency of node importance across repeated sampling. Nodes that consistently exhibit high centrality are interpreted as structurally important features of the CMD network.


    
    
    
    
    


\section{Results}\label{sec:Result}





\begin{figure}
    \centering
   \includegraphics[width=0.8\textwidth]{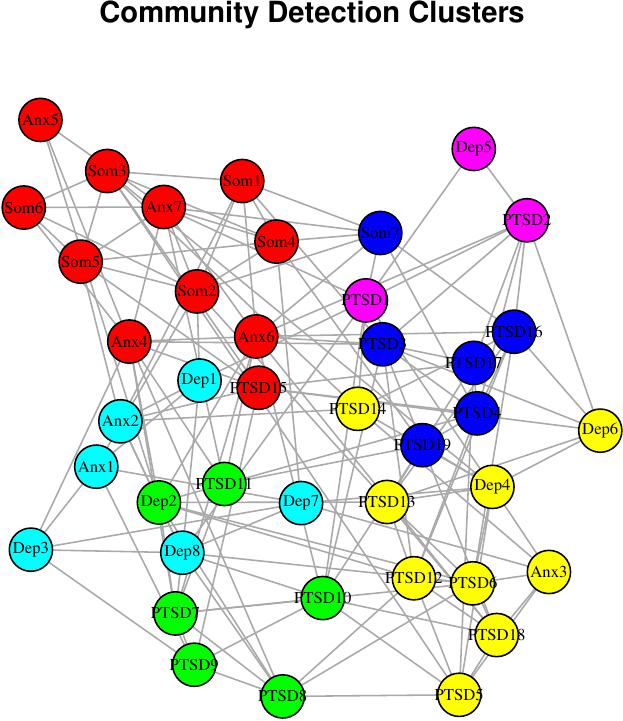}
   \caption{Network visualization with community-based coloring, where each color represents a community identified by the Walktrap algorithm implemented on estimated network structure. Nodes represent CMD symptoms across PTSD, depression, anxiety, and somatic domains. The somatic-anxiety cluster is colored red, the mixed emotional distress cluster is yellow, the PTSD-depression cluster is green, the depression–anxiety cluster is turquoise blue, the PTSD–somatic cluster is blue, and the migration-related distress cluster is purple.}
\label{fig:network_comparison}
\end{figure}

The community detection analysis based on the estimated network identified six distinct symptom clusters (communities) of CMD factors, revealing important patterns of symptom co-occurrence and potential shared mechanisms. These symptom clusters are highlighted in Figure \ref{fig:network_comparison}. Notably, none of the six identified clusters were composed exclusively of depressive, PTSD, somatic, or anxiety symptoms. Rather, symptoms of depression and anxiety consistently co-occurred with symptoms from other domains, particularly PTSD, showing the interconnected nature of CMD symptoms in this population. While we interpret the clusters in terms of potential dependence among symptoms, our aim is to identify patterns of co-occurrence and study associations rather than to infer causality or directionality.

The first cluster, the \textbf{Somatic Anxiety Cluster} (colored red in Figure \ref{fig:network_comparison}), includes a combination of somatic symptoms (Som1 to Som6), anxiety symptoms (Anx4 to Anx7), and PTSD15, reflecting a strong co-occurrence of physical complaints and anxiety-related experiences. This cluster suggests that physical symptoms such as dizziness, chest pain, nausea, difficulty breathing, cold spells, and tingling in the body frequently overlap with anxiety symptoms like feeling tense, feeling scared, feeling fearful, and experiencing a sense of terror or panic, which could in turn lead to sleep problems \citep{Lies2021}.


The \textbf{Mixed Emotional Distress Cluster} (colored yellow in Figure \ref{fig:network_comparison}) includes a mix of PTSD symptoms (PTSD5, PTSD6, PTSD12, PTSD13, PTSD14, PTSD18), along with two depressive symptoms (Dep4, Dep6) and a single anxiety symptom (Anx3). This cluster reflects the intersection of emotional and physical trauma responses. Here, we see that lack of interest in activities and feelings of worthlessness co-occur with anxiety symptoms such as feeling nervous and PTSD symptoms such as reliving past experiences, feeling very upset, feeling cut off from others, emotional numbness, a sense that one’s future may be cut short, and hypervigilance. The presence of feeling cut off from others further highlights the emotional detachment often experienced in trauma, particularly when it co-occurs with symptoms like feeling worthless and reliving past experiences \citep{Boehme2025, Kratzer2024, Arel2017, Gershuny1999}.

Next, PTSD symptoms such as PTSD7, PTSD8, PTSD9, PTSD10 and PTSD11, and a single depressive symptom (Dep2) are clustered in the \textbf{PTSD-Depression Cluster} (colored green in Figure \ref{fig:network_comparison}), reflecting a core grouping of trauma-related responses. This cluster shows that trauma such as having strong physical reactions could potentially lead an individual to avoid talking about stressful past events, avoid activities that remind them of those experiences, and lose interest in things they once enjoyed. In addition, since PTSD-related symptoms revolve around painful past experiences, the inclusion of post-migration depression in this cluster highlights the lingering psychological impact of that trauma on the Eritrean population.

The fourth cluster, the \textbf{Depression–Anxiety Cluster} (colored turquoise blue in Figure \ref{fig:network_comparison}) includes a combination of anxiety symptoms (Anx1, Anx2) and depressive symptoms (Dep1, Dep3, Dep7, Dep8), reflecting a blend of emotional distress that spans both pre- and post-migration experiences. This cluster shows that early anxiety, both before migration and after arriving in the United States, coexist with depressive symptoms such as feeling blue, feeling lonely, and experiencing suicidal thoughts. The presence of depression prior to migration alongside these symptoms suggests a possible pattern of sustained psychological burden that cuts across different phases of the refugee experience.

The next cluster, the \textbf{PTSD–Somatic Cluster} (colored blue in Figure \ref{fig:network_comparison}) includes the remaining PTSD symptoms (PTSD3, PTSD4, PTSD16, PTSD17, PTSD19) not captured in previous clusters, along with a single somatic symptom (Som7). This cluster reflects the intersection of psychological trauma and physical exhaustion. Here, weakness in the body appears alongside PTSD symptoms such as disturbing memories, disturbing dreams, feeling irritable, trouble concentrating, and feeling jumpy, all of which point to the potentially disruptive impact of trauma on day-to-day functioning \citep{harnett2022structural, thompson2010mindfulness}. 

Finally, the \textbf{Migration-Related Distress Cluster} (colored purple in Figure \ref{fig:network_comparison}) is a mix of the final two PTSD symptoms (PTSD1, PTSD2) and the remaining depressive symptom (Dep5). This cluster possibly reflects the emotional toll of migration, where trauma experienced both before and after arriving in the United States is grouped with feelings of hopelessness.

\subsection{Node Centrality Measures Analysis}\label{central}
In this subsection, we examine the CMD symptom network using three centrality measures as described before. To maintain interpretability and visual consistency, we continue with the community-based color scheme used in the cluster analysis results.

\begin{figure}[]
    \centering
    \includegraphics[width=0.8\textwidth]{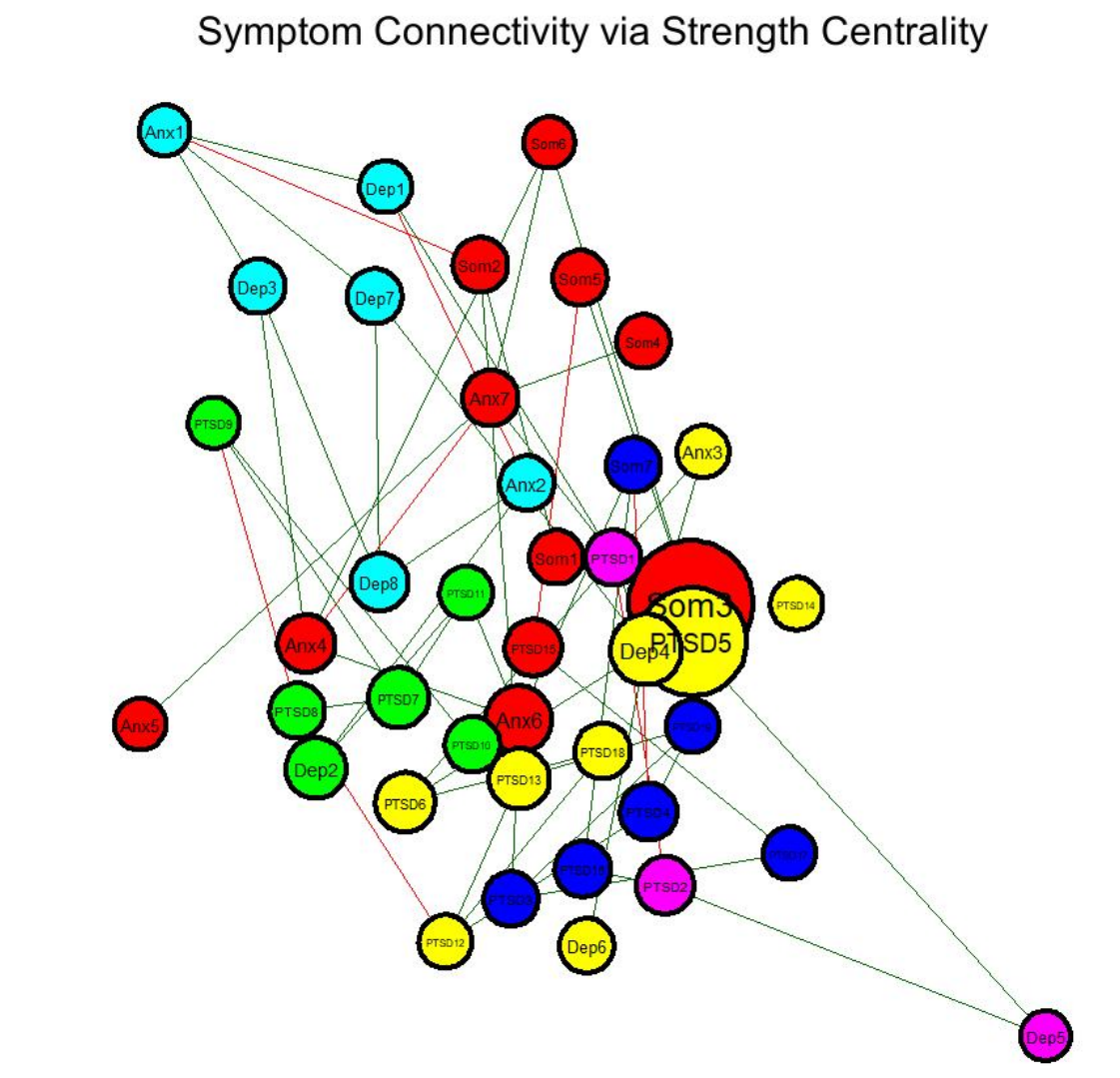} 
    \caption{This plot visualizes the strength centrality of CMD symptoms. Larger nodes reflect higher overall strength, based on the sum of their connection weights, while smaller nodes may still have a few strong edges but lower overall connectivity. Green lines represent positive associations, and red lines indicate negative associations. Colors follow the same community-based grouping shown earlier: red for the somatic-anxiety cluster, yellow for mixed emotional distress, green for PTSD-depression, turquoise blue for depression–anxiety, blue for PTSD–somatic, and purple for migration-related distress.}
    \label{fig:strengthness}
\end{figure}

The strength centrality measure (see Figure \ref{fig:strengthness}) identify nausea (Som3) and reliving past experiences (PTSD5) as the most central symptoms, with lack of interest in activities (Dep4) and feeling fearful (Anx6) also showing moderate importance. Nausea (Som3) connects strongly with hot or cold spells (Som5), weakness in the body (Som7), and nervousness (Anx3), positioning it as a hub in the somatic-anxiety cluster. Reliving past experiences (PTSD5), a key node in the mixed emotional distress cluster, links positively with hopelessness (Dep5). Though physical reactions to reminders (PTSD7) has lower overall strength, it still shows strong co-occurrence with related PTSD symptoms.

\begin{figure}[h!]
    \centering
    \includegraphics[width=0.8\textwidth]{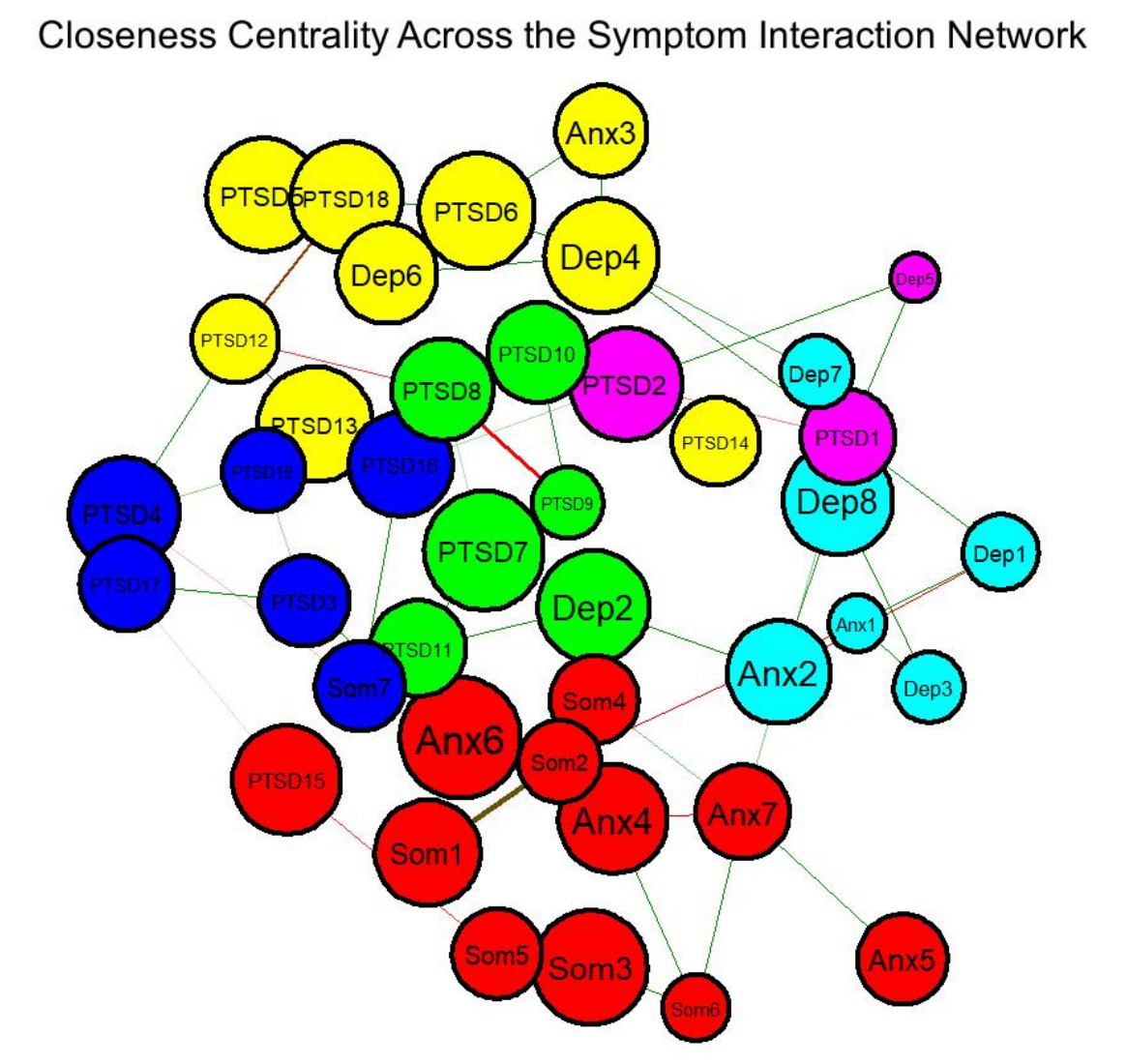} 
    \caption{
        \textbf{Closeness Centrality Plot.} This figure illustrates the closeness centrality of CMD symptoms in the network, highlighting nodes with the highest connectivity and influence across domains. Colors follow the same community-based grouping shown earlier: red for the somatic-anxiety cluster, yellow for mixed emotional distress, green for PTSD-depression, turquoise blue for depression–anxiety, blue for PTSD–somatic, and purple for migration-related distress.}
    \label{fig:closeness}
\end{figure}

The closeness centrality plot (Figure~\ref{fig:closeness}) reveals a wider range of structurally central symptoms, including reliving past experiences (PTSD5), hypervigilance (PTSD18), feeling very upset (PTSD6), lack of interest in activities (Dep4), emotional numbness (PTSD13), trouble concentrating (PTSD2), flashbacks (PTSD4), sense of a shortened future (PTSD16), loss of interest in things once enjoyed (Dep8), depression post-migration (Dep2), worrying too much (Anx2), feeling fearful (Anx6), feeling tense (Anx4), and nausea (Som3). These symptoms, present across all six clusters, are positioned to efficiently interact with the rest of the network despite varying degrees of direct connectivity.

\begin{figure}[h!]
    \centering
    \includegraphics[width=0.8\textwidth]{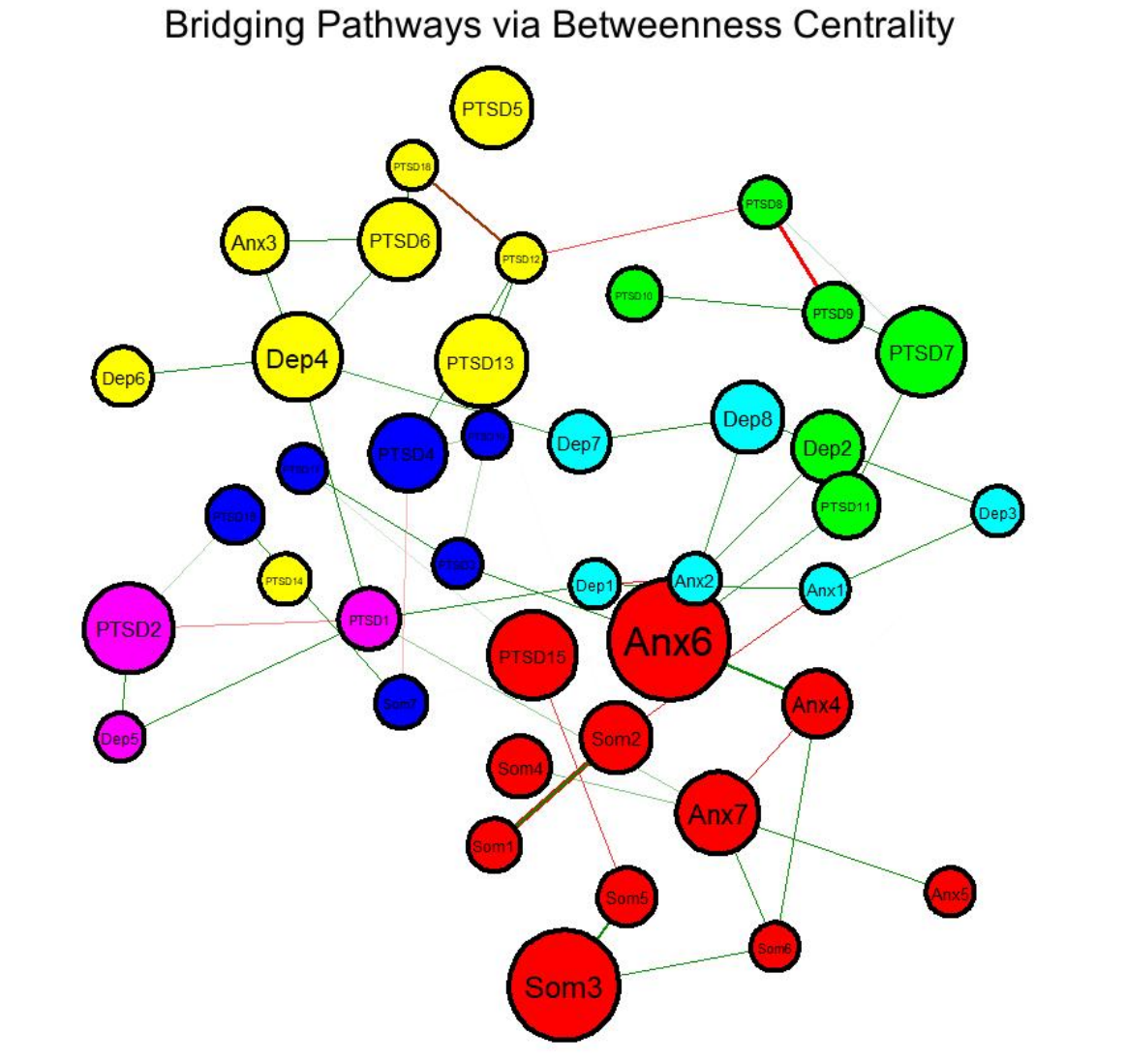} 
    \caption{\textbf{Betweenness Centrality Plot.} 
       This figure depicts the betweenness centrality of CMD symptoms, highlighting key nodes that act as bridges between symptom clusters. Colors follow the same community-based grouping shown earlier: red for the somatic-anxiety cluster, yellow for mixed emotional distress, green for PTSD-depression, turquoise blue for depression–anxiety, blue for PTSD–somatic, and purple for migration-related distress.}\label{fig:betweenness}
\end{figure}

The betweenness centrality plot (Figure~\ref{fig:betweenness}) highlights feeling fearful (Anx6) and nausea (Som3) as prominent bridge symptoms—suggesting they frequently lie on the shortest paths between other symptom pairs. Moderately central symptoms included reliving past experiences (PTSD5), feeling very upset (PTSD6), emotional numbness (PTSD13), trouble concentrating (PTSD2), flashbacks (PTSD4), physical reactions to reminders (PTSD7), sleep problems (PTSD15), sense of terror or panic (Anx7), and lack of interest in activities (Dep4). For example, feeling fearful (Anx6) connects to the depression–anxiety cluster via thoughts of ending life (Dep8), to the PTSD–somatic cluster through PTSD level post-migration (PTSD2), and to the PTSD–depression cluster through Depression post-migration (Dep2) and loss of interest in things once enjoyed (PTSD11). Meanwhile, Dep4 acts as a bridge to the migration-related distress cluster via PTSD level before migration (PTSD1).

Notably, nausea (Som3) and feeling fearful (Anx6) emerge as highly central symptoms in both strength and betweenness analyses, and show moderate centrality in closeness. This suggests that these symptoms not only maintain strong connections with others but also serve as crucial bridges between symptom clusters, facilitating the spread or interaction of distress across domains such as somatic, anxiety, and depressive symptoms. Meanwhile, reliving past experiences (PTSD5) and lack of interest in activities (Dep4) consistently rank high across all three centrality measures, indicating that they are both well-connected, structurally close to other symptoms, and frequently lie on the shortest paths within the network. Together, these overlaps suggest that interventions targeting Som3, Anx6, PTSD5, and Dep4 could possibly have a broad and cascading impact across the symptom network either by weakening key pathways of co-occurrence or by reinforcing recovery through central, widely connected symptom nodes. These symptoms may serve as effective leverage points for interrupting or alleviating widespread CMD symptom burden.

\subsection{Assessing Network Stability via Bootstrap Analysis}

\begin{figure}[ht!]
 \begin{subfigure}[t]{0.4\textwidth}
  \centering
     \includegraphics[width=\textwidth]{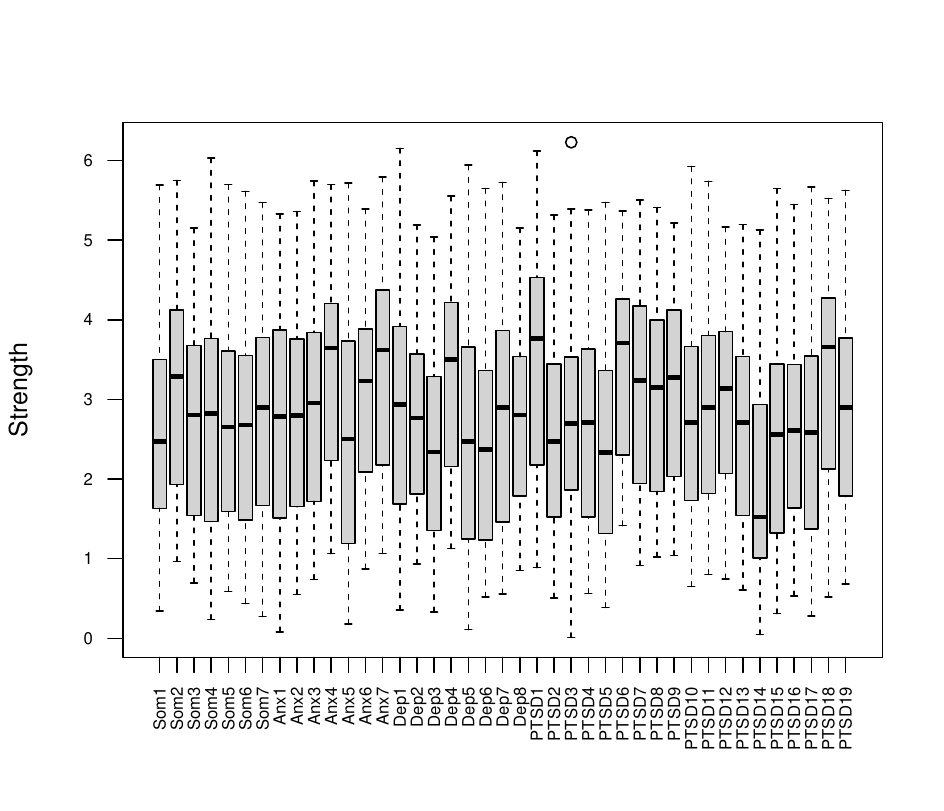}
     \caption{Strength Centrality}
     \label{fig:stren}
 \end{subfigure}
 \hfill
 \begin{subfigure}[t]{0.4\textwidth}
  \centering
     \includegraphics[width=\textwidth]{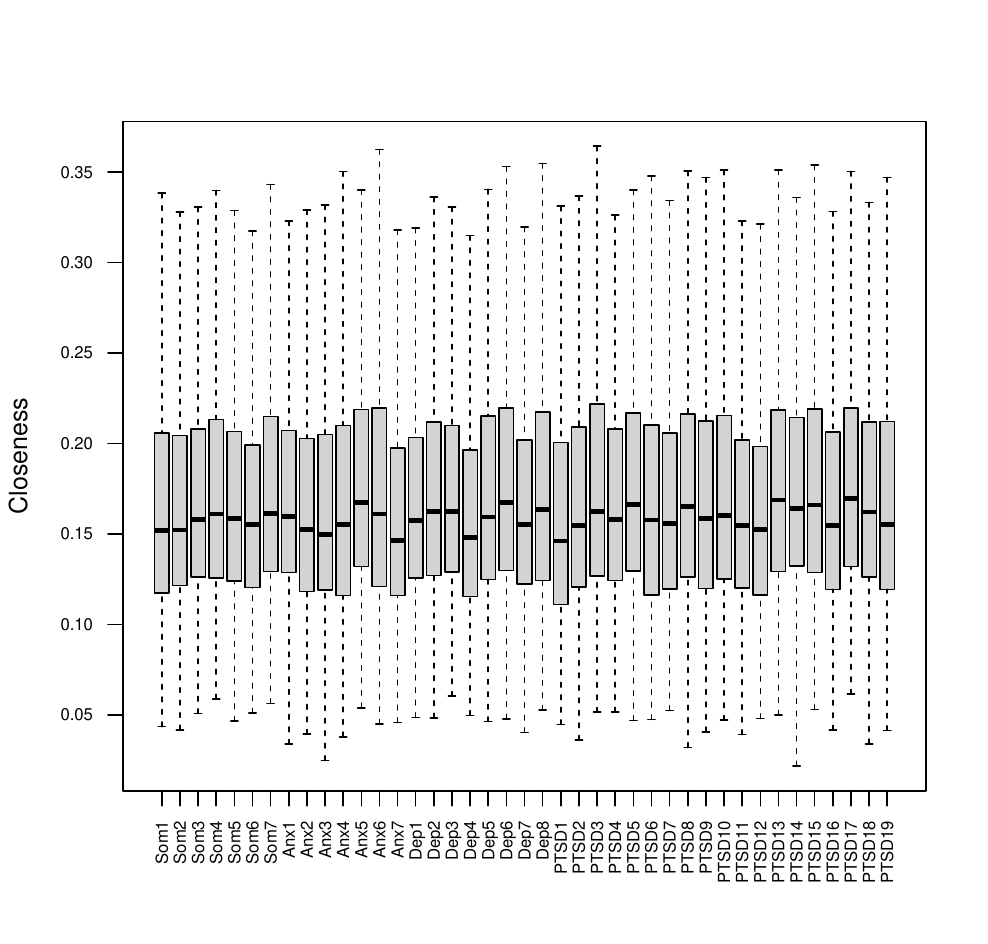}
     \caption{Closeness Centrality}
     \label{fig:close}
 \end{subfigure}
 \hspace*{\fill} 
 \begin{subfigure}[t]{0.4\textwidth} 
  \begin{center}
     \includegraphics[width=\textwidth]{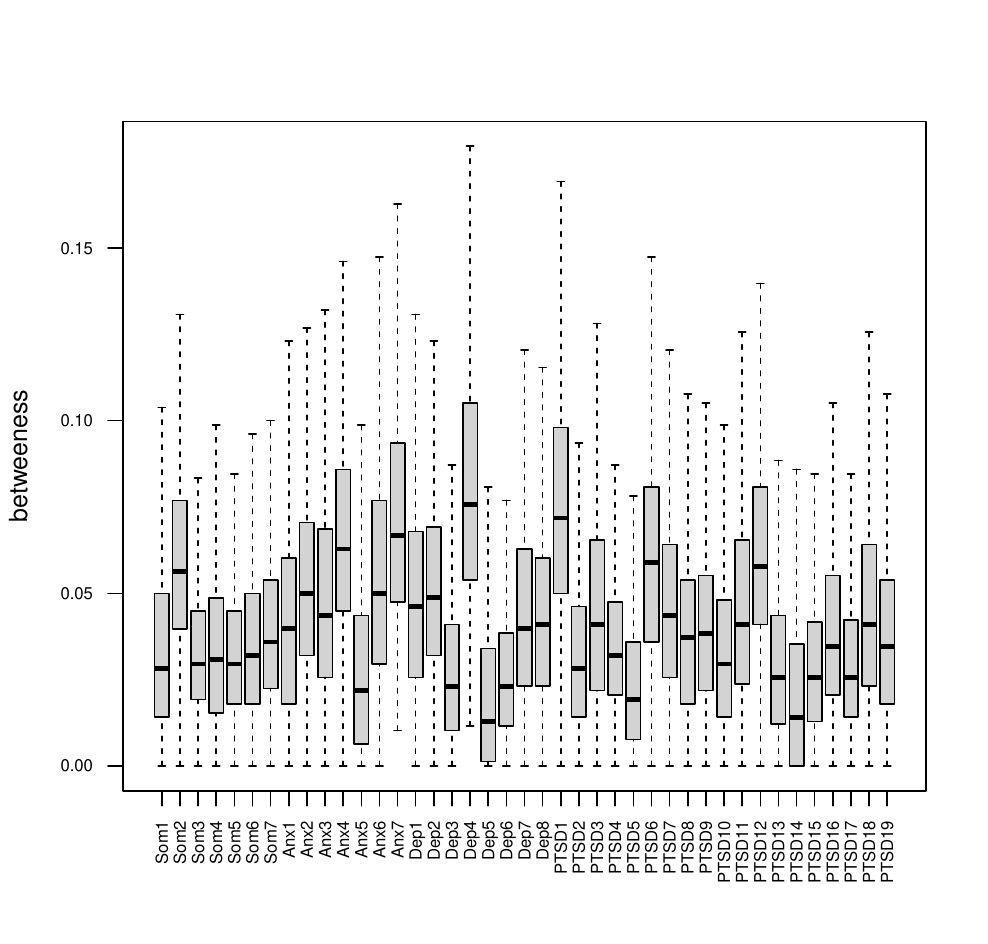}
     \caption{Betweenness Centrality}
     \label{fig:betw}
     \end{center}
 \end{subfigure}
   \hspace*{\fill} %
  \caption{Bootstrap distributions of centrality values for all nodes in the CMD network across 1,000 iterations, illustrated separately for (a) strength, (b) closeness, and (c) weighted betweenness. Each boxplot represents the variability in centrality estimates for a single node. As seen in (a) and (b), strength and closeness centrality remained relatively stable, with most nodes showing tightly clustered values across bootstraps. In contrast, (c) reflects greater variability in weighted betweenness centrality. While some nodes exhibited wider distributions, others appeared fairly stable—highlighting the global nature of this metric and its sensitivity to changes in network structure. Key nodes discussed in the results can be visually identified by their higher median values and narrower interquartile ranges.}\label{fig:boostrap}
\end{figure}

To evaluate the robustness of the network model, we performed a nonparametric bootstrap analysis based on 1,000 iterations. Figure \ref{fig:boostrap} presents boxplots for each centrality measure, strength, closeness, and betweenness, showing the distribution of centrality values for all 41 CMD symptoms across the resampled networks. These plots illustrate how centrality scores shifted across the bootstrap samples, making it easier to see which nodes consistently ranked high and which were more sensitive to changes in the data.

As shown in Figure~\ref{fig:boostrap}, strength and closeness centrality remained relatively stable, with moderate dispersion and consistent node rankings across iterations. Weighted betweenness centrality showed slightly broader variability across some nodes, but still demonstrated reasonable stability overall. For strength centrality, the most central nodes (such as, PTSD1, PTSD6, Anx7, Anx4) consistently ranked highest across all bootstrap samples, with their mean centrality values even exceeding the original estimates. The uniformity in mean differences (ranging from 0.95 to 1.3) and moderate standard deviations (0.97 to 1.38) underscore the reliability of the central roles of these nodes. Similarly, closeness centrality revealed a tightly clustered set of key symptoms (namely, PTSD15, Anx6, Dep2, PTSD5) with bootstrap means closely matching original values. The stability in rankings and uniformly low variability (SDs between 0.08 and 0.10) indicates these symptoms reliably maintained central positions across network iterations. For betweenness centrality, the use of weighted paths led to considerably improved consistency. Core nodes such as Dep4, Anx7, and PTSD1 showed only minor deviations between original and bootstrap values (typically < 0.05), and their standard deviations remained low (0.02–0.04), affirming their robustness across resampled networks.

This comparison highlights that our centrality results are not artifacts of random sampling variation but reflect a stable and reproducible structure in the symptom network. The key nodes consistently emerge as central across multiple resampled networks, reinforcing confidence in their importance for future possible targeted interventions.

\section{Discussion}\label{sec:Discuss}
Analyzing symptom networks in high-dimensional settings introduces unique statistical challenges, especially when the number of symptoms ($p$) far exceeds the available sample size ($n$). In this study, we addressed the small sample problem ($n < p$) by extending the graphical lasso framework through adaptive regularization, resulting in what we refer to as adaptive glasso. This regularized approach enhances both sparsity selection and stability in high-dimensional network estimation. Traditional covariance-based methods often produce singular or unstable precision matrices under such conditions, making inverse estimation unreliable. Our novel adaptive glasso addresses this by applying differential shrinkage, preserving important symptom relationships while minimizing the risk of overfitting.

The symptom clusters in our network revealed six distinct communities, each representing meaningful symptom co-occurrence patterns across PTSD, depression, anxiety, and somatic domains. These clusters show interactions between psychological and physiological distress (see Figure~\ref{fig:network_comparison}), such as somatic complaints and anxiety-related symptoms (somatic-anxiety cluster) and relationships between PTSD symptoms and emotional detachment (mixed emotional distress cluster). We also examine node centrality measures to identify key symptoms that serve as core drivers of symptom interactions and connectivity within the network.

In the strength centrality plot (Figure~\ref{fig:strengthness}), nausea (Som3) and reliving past experiences (PTSD5) stand out as the main central hubs, suggesting that they have an overall positive relationship with their neighboring symptoms. This means these two central hubs, in terms of connectivity, could intensify together with other symptoms or ease together. Meanwhile, loss of interest in activities (Dep4) and feeling fearful (Anx6) show moderate overall connectivity with their neighbors, so these two symptoms should not be overlooked in the network. Thus, based on strength centrality, Som3, PTSD5, Dep4, and Anx6 emerge as important symptoms that could serve as promising targets for intervention, given their overall connection to other symptoms in the network.

In terms of closeness centrality (Figure~\ref{fig:closeness}), symptoms such as PTSD5, PTSD18, PTSD6, Dep4, PTSD13, PTSD2, PTSD4, PTSD16, Dep8, Anx2, PTSD7, Dep2, Anx6, Som3, and Anx4 appeared moderately larger than others. These symptoms could efficiently reach many other symptoms across the network. Using this knowledge, targeting groups of these symptoms could result in more effective support of the Eritrean population due to the high closeness of these symptoms between the clusters.

In the betweenness centrality plot (Figure~\ref{fig:betweenness}), feeling fearful (Anx6), nausea (Som3), and possibly loss of interest in activities (Dep4), PTSD level post-migration (PTSD2), and physical negative reaction (PTSD7) serve as critical nodes that help keep different parts of the network connected. Based on betweenness centrality, these symptoms could be particularly valuable intervention targets, as they bridge across different symptom clusters and help maintain the overall network structure.

Some studies have reported related outcomes, and it's interesting to see that our findings align with parts of their results, even though some of these studies did not specifically use centrality measures, but rather examined relationships between CMD symptoms within their contexts. For instance, \cite{Zhou2020} identified a strong relationship between lack of interest in things (Dep4 in our study) and depression, along with other key links involving avoidance of thoughts (PTSD8), hypervigilance (PTSD18), and exaggerated startle response (PTSD19). Supporting this, a PTSD and depression network analysis among firefighters by \cite{Cheng2023} found that loss of interest in things (PTSD11) was a highly central symptom, alongside emotional numbness (PTSD13) and high alertness (PTSD18). Similarly, \cite{Bryant2017} reported that intrusive memories (PTSD3) and reliving past activities/experiences (PTSD5) were particularly central in the acute phase of PTSD. Their findings suggest that early intervention on either symptom could help prevent broader PTSD symptom progression—consistent with our finding that PTSD5 emerged as one of the most central symptoms in our network. Another study by \cite{Duek2021} identified feeling distant or cut off from others (PTSD12) and disturbing memories (PTSD3) as central symptoms. Although these symptoms did not emerge as central hubs or critical nodes in our network, they showed moderate prominence in closeness centrality, suggesting they still play important roles.

Beyond studies highlighting specific symptoms, several researchers have focused more directly on refugee populations and sleep disturbances consistently emerge as major concerns. For example, \cite{Lies2021} found that sleep disturbances (PTSD15) mediated the relationship between trauma exposure and posttraumatic stress symptoms among Syrian and Iraqi refugees. They suggested that targeting sleep disturbances could be an important part of interventions aimed at improving psychological outcomes. Similarly, \cite{Sankari2023} reported that PTSD-related sleep problems were strongly associated with PTSD symptoms and postmigration difficulties in Syrian refugees, reinforcing the importance of addressing sleep disturbances. In addition, \cite{AlSmadi2022} found that sleep disorders (especially insomnia) were highly prevalent among resettled Syrian refugees, playing a key role in linking psychological distress with somatic symptoms. Their findings show how sleep problems, such as difficulty staying asleep, can act as central hubs in symptom networks, which aligns closely with our observation that sleep disturbances (PTSD15) appeared moderately sized in both betweenness and closeness centrality. Overall, across all three centrality measures, symptoms such as feeling fearful (Anx6), nausea (Som3), loss of interest in activities (Dep4), and sleep problems (PTSD15) seem to stand out for the Eritrean population. These symptoms could serve as strong starting points when prioritizing intervention strategies, especially when focusing on a few key areas at a time to better support this community.

Building on these findings, it is important to consider how such symptoms should inform the design of interventions. Migrant health literature prioritizes migrant mental health due to the increased vulnerability of refugees and migrants to mental health challenges arising from migration-related stressors. Studies indicate that psychosocial interventions for migrants predominantly focus on mental health issues \citep{afuwape2010cares, arundell2021effectiveness, baskin2021community}. However, our study highlights the need to broaden the scope of interventions to include a holistic framework that addresses both physical and mental health. One of the significant observations emerging from our symptoms network analysis is the centrality of somatic symptoms, such as nausea, which are often culturally embedded expressions of distress. In many cultures, such as among Eritreans, somatic presentations may serve as the primary mode of expressing underlying psychological suffering, rather than signaling discrete physical ailments. As such, generic health interventions that do not account for cultural idioms of distress may fail to adequately address the needs of migrant populations. Understanding how different cultures manifest and express psychological and physical distress is essential for the development of effective, culturally responsive care. Culturally sensitive and competent healthcare practices are therefore critical for enhancing patient engagement and improving outcomes. By incorporating a deeper understanding of cultural factors and embracing a holistic approach to health, interventions for migrants and refugees can become more nuanced, effective, and aligned with the lived experiences of those they aim to serve.

\section{Conclusion}

Our analysis shows that unweighted betweenness centrality appeared less robust under bootstrap resampling compared to strength and closeness. This is likely due to the global, path-based nature of betweenness, which depends on the shortest paths between node pairs \citep{freeman1977set}. Even small changes in network structure can alter these paths, resulting in considerable variability in betweenness values. This aligns with prior findings that betweenness is among the least stable centrality measures \citep{duron2019variability}, especially when subjected to data resampling procedures \citep{kardos2020stability, costenbader2003stability, segarra2015stability}. Such instability can be especially pronounced in high-dimensional networks where the number of nodes far exceeds the number of available observations. To address this, we employed weighted betweenness alongside adaptive graphical regularization (adaptive glasso), which promotes sparsity and improves the stability of network estimation. Our novel application of adaptive penalties in combination with weighted betweenness helped reduce the observed variability and improved reliability in betweenness estimates.

Although shortest-path calculations are generally less reliable in high-dimensional, undirected graph where stable alternative paths are limited \citep{kardos2020stability}, our approach, which defines edge weights as the inverse of the absolute values in the precision matrix, provides a more stable foundation for measuring global centrality. As a result, both locally oriented measures (strength and closeness) and the global path-based measure (weighted betweenness) demonstrates relative stability under resampling, supporting the overall robustness of our network model.


A key limitation of this study is that, while centrality analysis identifies structurally important symptoms within the network, it does not reflect causal pathways and should not be interpreted as evidence of symptom progression at the individual level. In addition, while centrality measures provide valuable insights into node importance, they are known to be sensitive to factors such as sample size, edge estimation methods, and model assumptions. Prior research has raised concerns about the stability of these metrics, particularly in psychometric networks estimated under high-dimensional constraints, where $n < p$ \citep{robinaugh2016identifying, epskamp2018estimating}. In light of these concerns, we interpret our centrality findings with appropriate caution. Nonetheless, the core contribution of this work lies in addressing the $n < p$ problem within a network modeling framework—an advance we believe is both timely and relevant to psychological and clinical network research.

\bibliographystyle{apalike}

\bibliography{reference}
\end{document}